\documentclass[iicol, sn-nature]{sn-jnl}

\usepackage{graphicx}%
\usepackage{multirow}%
\usepackage{amsmath,amssymb,amsfonts}%
\usepackage{amsthm}%
\usepackage{mathrsfs}%
\usepackage[title]{appendix}%
\usepackage{xcolor}%
\usepackage{textcomp}%
\usepackage{manyfoot}%
\usepackage{booktabs}%
\usepackage{algorithm}%
\usepackage{algorithmicx}%
\usepackage{algpseudocode}%
\usepackage{listings}%
\usepackage{subcaption}
\usepackage{enumitem}

\begin{document}

\title{Machine Learning Based Efficiency Calculator (MaLBEC) for Nuclear Fusion Diagnostics}

\author*[1,2]{\fnm{Kimberley} \sur{Lennon}}\email{kimberley.lennon@ukaea.uk}

\author[2]{\fnm{Chantal} \sur{Shand}}\email{chantal.shand@ukaea.uk}

\author[2]{\fnm{Gemma} \sur{Wilson}}\email{gemma.wilson@ukaea.uk}

\author[1]{\fnm{Robin} \sur{Smith}}\email{robin.smith@shu.ac.uk}

\affil[1]{\orgdiv{School of Engineering and Built Environment}, \orgname{Sheffield Hallam University}, \orgaddress{\street{Howard Street}, \city{Sheffield}, \postcode{S1 1WB}, \country{UK}}}

\affil[2]{\orgdiv{Applied Radiation Technology Group}, \orgname{UK Atomic Energy Authority}, \orgaddress{\street{Culham Campus}, \city{Abingdon}, \postcode{OX14 3DB}, \country{UK}}}


\abstract{
Diagnostics are critical for commercial and research fusion machines, since measuring and understanding plasma features is important to sustaining fusion reactions. The neutron flux (and therefore fusion power) can be indirectly calculated using neutron activation analyses, where potentially large numbers of activation foils are placed in the neutron flux, and delayed gammas from key reactions are measured via gamma spectrometry. In gamma spectrometry, absolute efficiency forms part of the activity calculation, and equals to the ratio of the total number of photons detected to the number emitted by a radioactive sample. Hence, it is imperative that they are calculated efficiently and accurately. This paper presents a novel digital efficiency calculation algorithm, the Machine Learning Based Efficiency Calculator (MaLBEC), that uses state-of-the-art supervised machine learning techniques to calculate efficiency values of a given sample, from only four inputs. In this paper, the performance of the MaLBEC is demonstrated with a fusion sample and compares the values to a traditional efficiency calculation method, Monte Carlo N-Particle (MCNP). The efficiencies from the MaLBEC were within an average 5\% of the ones produced by MCNP, but with an exceptional reduction in computation time of 99.96\%. When the efficiency values from both methods were used in the activity calculation, the MaLBEC was within 3\% of the MCNP results.
}


\maketitle

\section{Introduction}\label{sec:intro}

Gamma spectrometry is a method in experimental nuclear physics used for identifying and quantifying photon radiation, by measuring $\gamma$ rays emitted from activated materials and naturally occurring background radiation. In fusion, gamma spectrometry is used in waste characterisation, materials research for future fusion machines, plasma diagnostics (e.g. indirectly calculating the neutron flux, and therefore fusion power, via neutron activation analysis \cite{MAST_U_indiumfoil,Lees_paper}), and other areas. High purity germanium (HPGe) detectors are often selected to measure low intensity or complex $\gamma$ ray signatures due to their excellent $\sim$~keV spectral resolution \cite{MLCSA_KL} and will be the focus of this paper. Once a spectrum has been collected with an HPGe detector, the activity, which is the number of counts per second observed by the detector, is derived from the energy spectrum, and is calculated for each photopeak as 

    \begin{equation} \label{eq1}
      a = \frac{A_n}{\varepsilon \times t \times b_r}, 
    \end{equation}
where $a$ is activity (Bq), $A_n$ is net peak area, $\varepsilon$ is the absolute efficiency, $t$ is the live counting time, and $b_r$ is the branching ratio. 

\begin{figure}[h]
    \centering
    \includegraphics[width=0.5\textwidth]{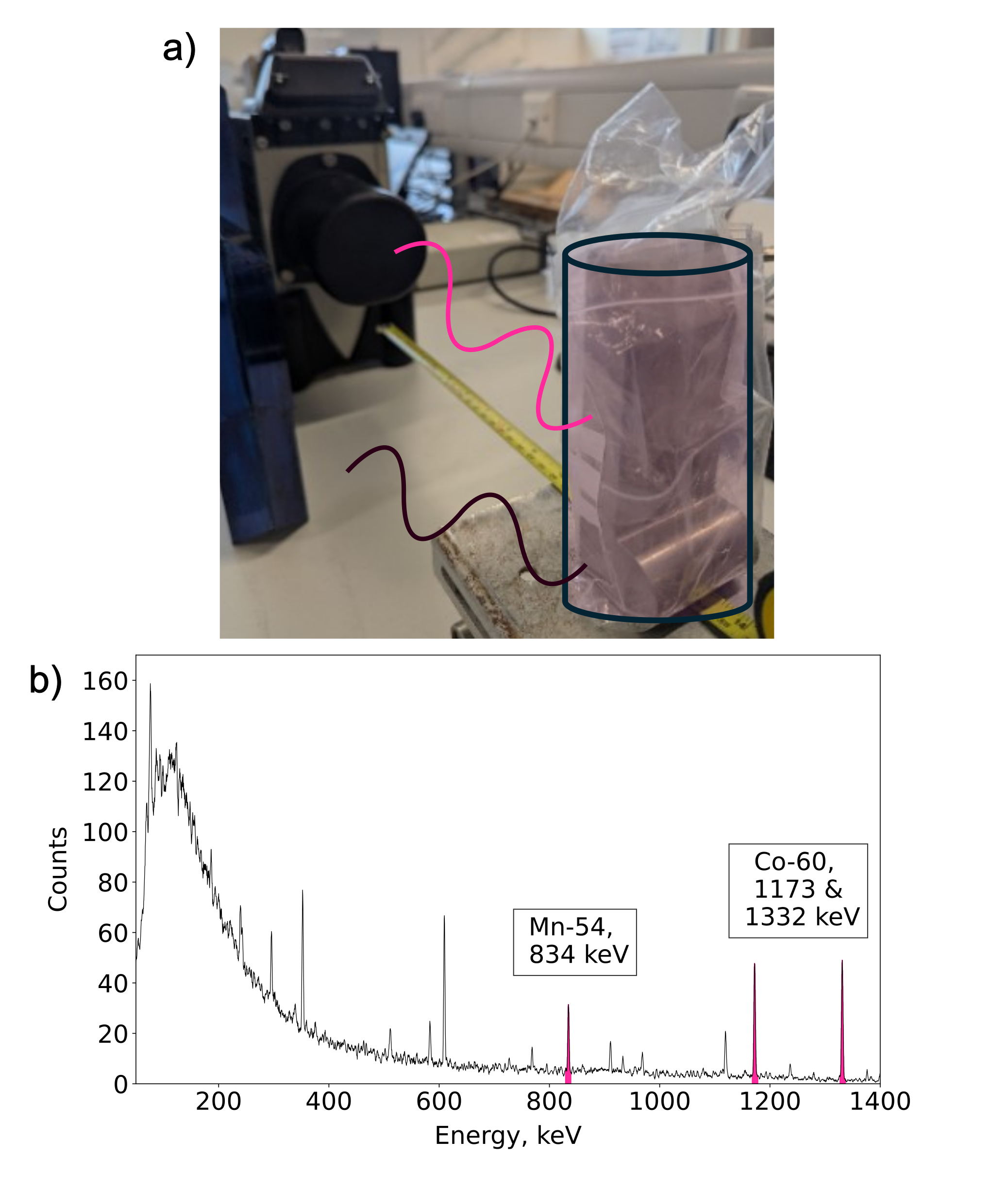}
    \caption{a) Experimental measurement set-up, including the Trans-SPEC-DX-100T HPGe detector and a steel sample (2~cm radius, 13.5~cm height, and a density of 6.89~$\mathrm{g/cm^3}$ positioned 30~cm from the detector). Radiation emitted from the sample was measured for 800~seconds. To demonstrate the importance of the efficiency calculation, the pink (light) wavy line demonstrates a photon that reaches the detector and the black (dark) wavy line is one that is not incident on the detector and so will not be seen.  b) Spectrum from the set-up shown in a), with the key photopeaks, manganese-54 and cobalt-60, labelled and filled in pink and the rest of the spectrum outlined in black}
    \label{fig:efficiency}
\end{figure}

The absolute efficiency ($\varepsilon$) is the ratio of the total number of photons detected to the number emitted by a radioactive sample, and is dependent on the sample geometry and density, detector geometry, photon energy, photon interactions, and the sample-to-detector position \cite{practical_g_spec}. The efficiency values form a key part of the activity calculation, hence it is imperative that they are calculated effectively and accurately. Figure~\ref{fig:efficiency} shows an experimental set-up of a fusion-relevant steel sample and the spectrum collected. This sample and spectrum are used in section~\ref{results} to provide validation of this work.

Traditional methods for calculating efficiency values are effective, but often require expensive software, difficult to obtain licences, extensive training and expertise, significant time to create or run models, reference sources in the same geometry as each new sample, and/or restrictive software that is difficult to use in tandem with laboratory automation tools. 

This paper presents a new method for absolute efficiency calculations, the machine learning based efficiency calculator (MaLBEC), which calculates the absolute efficiencies for 11 pre-determined photon energies for any cylindrical sample based on just four inputs.

\section{Methods}\label{methods}

\subsection{Data collection and preprocessing}
The data used in selecting the machine learning model and training the MaLBEC were simulated via Monte Carlo N-Particle (MCNP). All calculations were performed using internal UK Atomic Energy Authority (UKAEA) Intel Xeon E5-2640 high-performance computing cluster with 16 CPU cores, two sockets per physical node, and 125~GB of RAM.

\begin{figure}[h]
    \centering
    \includegraphics[width=0.5\textwidth]{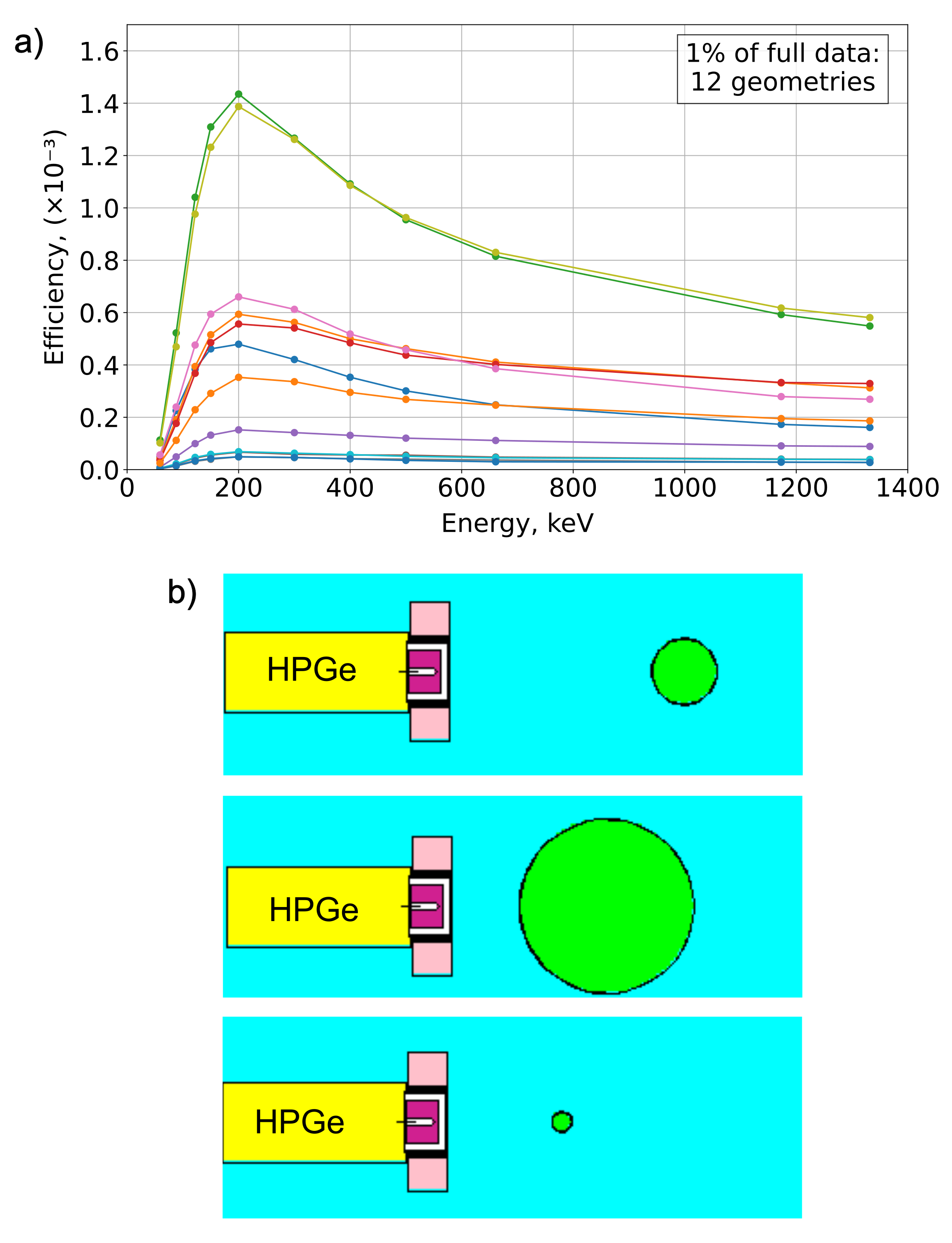}
    \caption{a) Efficiency curves of 1\% of the full 1258 simulated cylindrical geometries. b) Top down view from MCNP of 3 of the 1258 geometries simulated, showing example variations in the simulations, such as varying radius and detector-to-sample distance. The detector is on the left and the cylindrical samples are on the right}
    \label{fig:data}
\end{figure}

The detector, modelled in MCNP, was the Trans-SPEC-DX-100T HPGe detector (as shown in figure~\ref{fig:efficiency}), which is housed in the radiological assay and detection lab (RADLab) at UKAEA. A total of 1258 MCNP models were generated by randomly varying four parameters of cylindrical items within set limits, which were determined to mimic likely sample geometries and measurements in the RADLab. The four parameters were sample density, sample height, sample radius, and distance to the detector. Each geometry required 11 MCNP files, one per output energy, in order to generate an efficiency curve. The 11 gamma ray energies were chosen to include the nuclides most relevant in fusion measurements and to cover the full spectrum energy range of interest to enable interpolation between energies. These were: 59~keV, 88~keV, 122~keV, 150~keV, 200~keV, 300~keV, 400~keV, 500~keV, 661~keV, 1173~keV, and 1332~keV. The full data set (1258 MCNP models), was split into 80\% training data (1006 geometries) and 20\% test data (252 geometries). Figure~\ref{fig:data} shows the efficiency curves of 1\% of the full data set (1\% so that the curves can be seen clearly without overlap) and figure~\ref{fig:data} shows some example MCNP models.

\subsection{Machine learning model selection and training}
The MaLBEC is a supervised regression machine learning model, specifically using a multilayer perceptron (MLP). These MLPs are a class of feedforward artificial neural networks (ANN) and they comprise an input layer, one or more hidden layers, and one output layer \cite{MLP_scikitlearn}.
The MLPs are trained using an algorithm called the backpropagation training algorithm, where the data are randomly shuffled and split into mini-batches (mini-batch size is fixed at 200 samples), which are passed through the MLP multiple times $-$ each full pass of all mini-batches is called an epoch \cite{ML_handson_book}. During training, the weights (the internal parameters that control the model’s predictions) are updated once per mini-batch — these updates are called iterations. On the forward pass, predictions are made and stored for each neuron (a computational unit that receives inputs, performs calculations, and produces an output \cite{goodfellow}), and then the output error is calculated using a loss function to compare the true output to the predicted output of the network \cite{ML_handson_book}. On the backward pass, the algorithm then computes how much each output connection contributed to the error, and then how much each error came from the connections in the layer below. This repeats until it reaches the input layer \cite{ML_handson_book}. Finally, the algorithm performs a gradient descent, which is an optimisation algorithm that adjusts all weights in the network to reduce the errors until the model converges on a minimum \cite{ML_handson_book}. The process is summarised below. 

\begin{enumerate}[label=\arabic*.]
    \item Epoch 1 (an epoch is a full run pass of all samples)
    
    \begin{enumerate}[label=\Alph*.]
        \item Randomly shuffle the data
        \item Split into mini-batches of 200 samples
            
            \begin{enumerate}[label=\roman*.]
            \item Iteration 1 (mini-batch 1)
                
                \begin{enumerate}[label=\alph*.]
                    \item Forward pass to predict output
                    \item Compute loss by comparing predictions to true values
                    \item Backpropagation to calculate the error at the output layer
                    \item Update weights to improve future predictions
                \end{enumerate}
            \item Repeat for each mini-batch / iteration
            
            \end{enumerate}
    \end{enumerate}

    \item Repeat for as many epochs as required until convergence or early stopping. 
\end{enumerate}

\begin{figure}[h]
    \centering
    \includegraphics[width=0.5\textwidth]{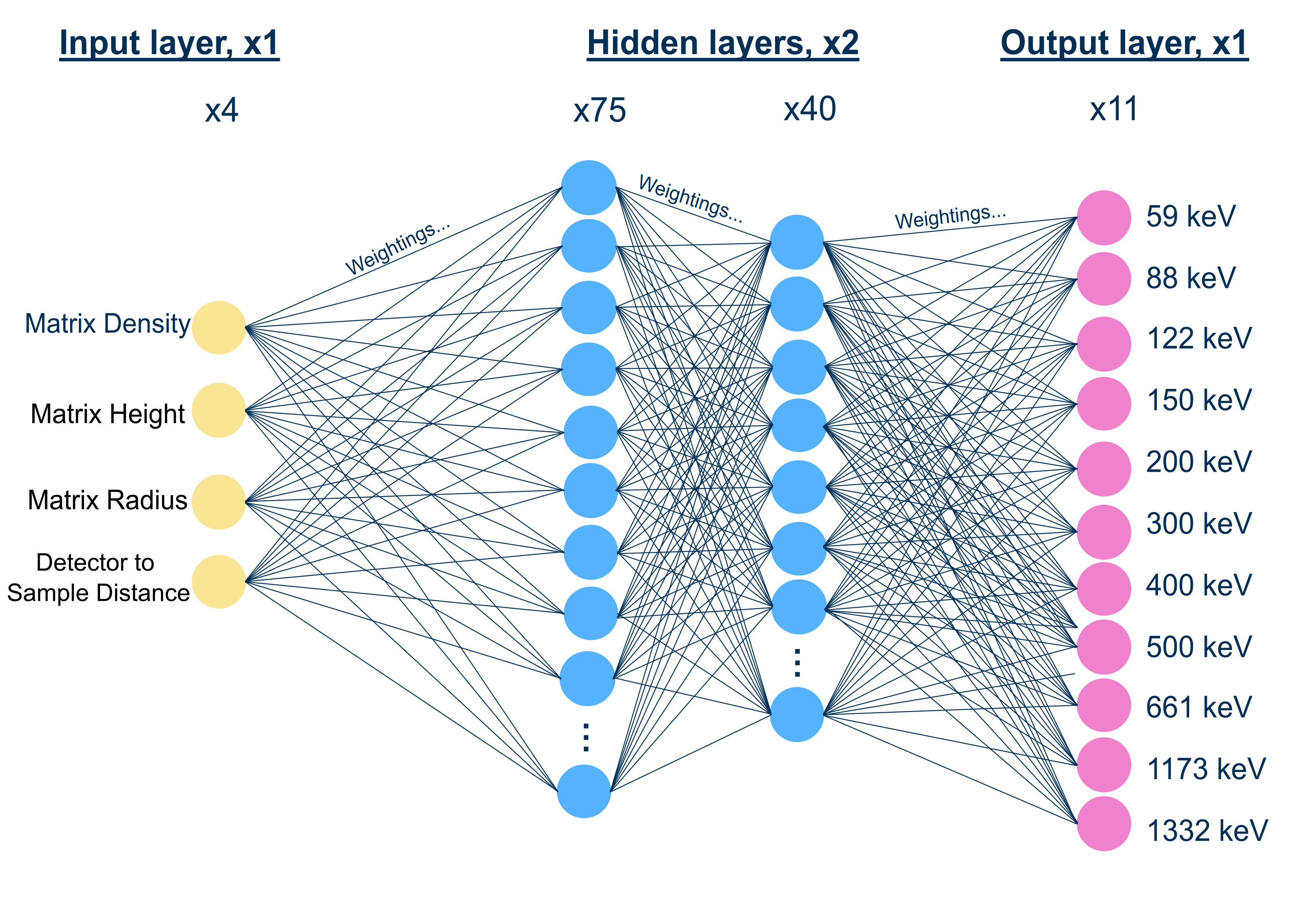}
    \caption{Architecture of the MLP used in the MaLBEC, showing the input layer with four inputs, two hidden layers, and an output layer with 11 neurons. Each layer, except the last layer, is fully connected to the next layer}
    \label{fig:malbec}
\end{figure}

The MLP architecture (shown in figure~\ref{fig:malbec}) was chosen via hyperparameter tuning through the grid search optimisation technique. The optimised architecture comprised an input layer with 4 neurons, 2 hidden layers size (75, 40), and an output layer with 11 neurons. The other tuned hyperparameters include the activation function ReLU, alpha (the learning rate which controls the step size during model training) of 0.12, initial learning rate of 0.01, max iterations of 10,000 (but early stopping was enabled with a validation fraction of 0.12 and number of iterations with no change set to 20), and the default solver of Adam. The architecture was implemented through standard \emph{sklearn} and \emph{keras} Python libraries.

\subsection{Machine learning model evaluation}
The performance of the trained MLP was evaluated through several metrics, including root mean squared error (RMSE) as shown in equation~\ref{eq:rmse} \cite{rmse} and cross-validation (CV) score \cite{ML_handson_book}. These metrics were applied on both the training and testing dataset to evaluate the model’s accuracy and generalisation capability. 

\begin{equation}
\text{RMSE} = \sqrt{ \frac{1}{n} \sum_{i=1}^{n} (y_p - y_a)^2 }
\label{eq:rmse}
\end{equation}

\section{Results}\label{results}

\subsection{Machine learning model results}
The RMSE was calculated for the training data, the test data, and for the mean CV score data, and the results are shown in table~\ref{tab:rmse_table}. To compare differences between the mean CV score and the training and testing data, the percentage difference equation was used,

    \begin{equation} \label{percent_dif}
      \% \textrm{dif} = \left ( \frac{a - b}{b} \right )  \times 100,
    \end{equation}
\noindent where $a$ and $b$ refer to the relevant value from the experimental value and actual/baseline value, respectively.

The results in table~\ref{tab:rmse_table} show that overall the RMSE values are very small (the efficiency values are approximately $1 \times 10^{-4}$, so an RMSE of less than $1 \times 10^{-5}$ is desirable as this is within 10\%). This low training RMSE suggests that the model has fit the training data well, but it still performs very well on unseen data, as shown by the low test RMSE. The training and testing RMSE are within the same order of magnitude, which shows that the model is not overfitting and generalises well to new data. The 38\% increase in RMSE between training and mean CV is expected and suggests that the model is not memorising the training data, and performs consistently across the 10 different ``folds'' $-$ subsets of the data used in CV to train and validate the model $-$ showing again that the model generalises well.

\begin{table}[h]
\caption{The RMSE results for the training, testing and mean CV datasets, with a percentage difference comparison.}\label{tab:rmse_table}%
\begin{tabular}{@{}ccccc@{}}
\toprule
\multicolumn{1}{c}{\begin{tabular}[c]{@{}c@{}}Training\\ RMSE\end{tabular}} &
\multicolumn{1}{c}{\begin{tabular}[c]{@{}c@{}}Mean CV\\ RMSE\end{tabular}} &
\multicolumn{1}{c}{\begin{tabular}[c]{@{}c@{}}Test\\ RMSE\end{tabular}} &
\multicolumn{1}{c}{\begin{tabular}[c]{@{}c@{}}\%dif CV\\ to Train\end{tabular}} &
\multicolumn{1}{c}{\begin{tabular}[c]{@{}c@{}}\%dif CV\\ to Test\end{tabular}} \\
\midrule
1.17e-05 & 1.61e-05 & 2.30e-05 & 38 & 30  \\
\botrule
\end{tabular}
\end{table}

Another indication that the model is not overfitting is that the early stopping was triggered at 159 iterations, despite a maximum iteration of 10,000 being available. This shows that the model converged quickly, preventing overfitting and over-optimising. Along with early stopping, the alpha value was tuned during the hyperparameter grid search to find the balance between overfitting and underfitting. This resulted in a strong alpha value being used, to prevent overfitting by penalising complexity \cite{ML_handson_book}. 

\begin{figure}[h]
    \centering
    \includegraphics[width=0.48\textwidth]{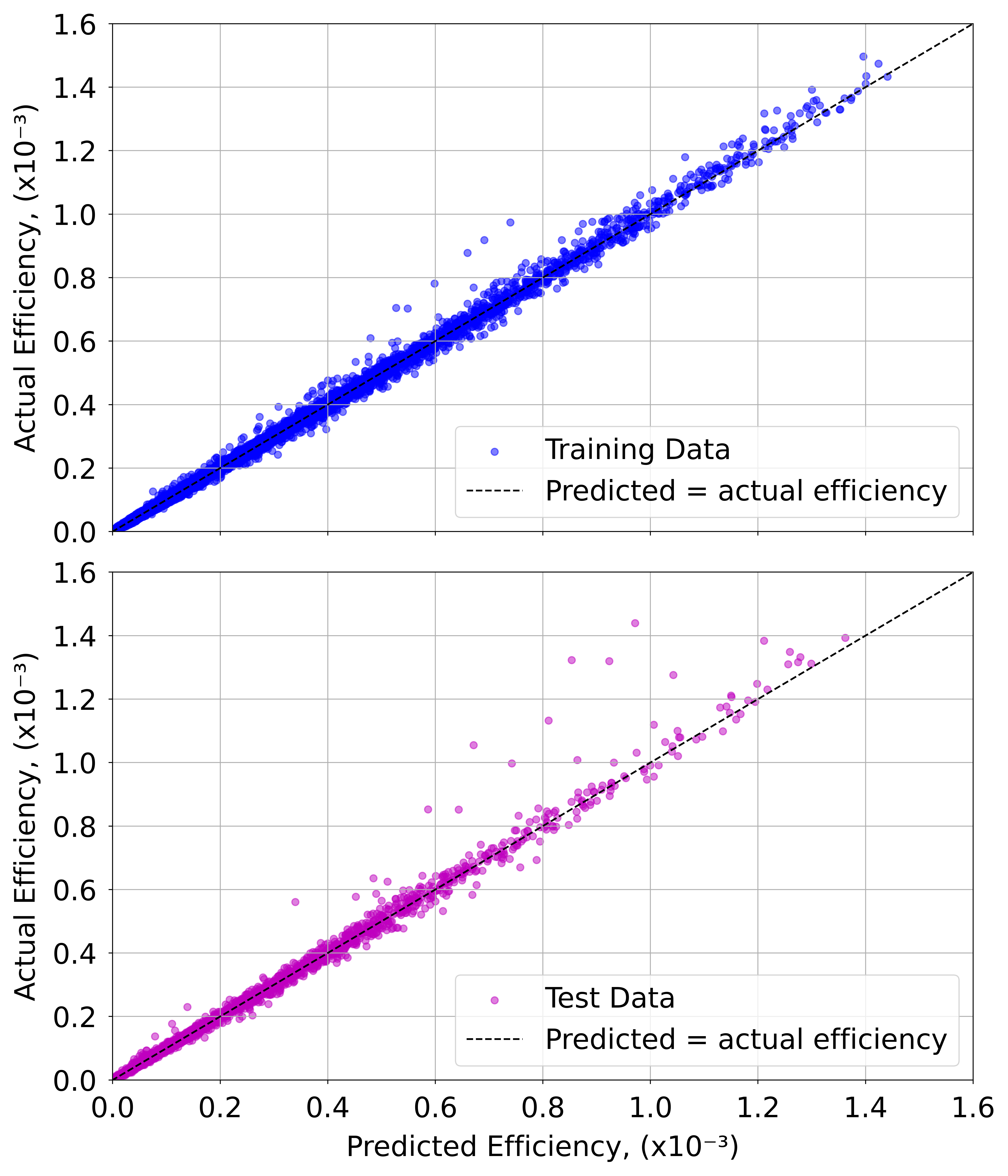}
    \caption{Actual vs predicted efficiency values for the training data (top/blue) and test data (bottom/pink)}
    \label{fig:model_results}
\end{figure}

A final indication that the model performs well is the comparison of the actual efficiency to predicted efficiency, on the training and test data. This comparison is shown in figure~\ref{fig:model_results}, where the plots have a similar pattern. The few outliers on the test data relate to the different geometries in the available data sets, where the training data didn't span the full parameter space due to limitations in the random generation of sample geometries. Despite this, the model performed well and could generalise to new, unseen geometries. 

\subsection{Comparative analysis with an experimental measurement}\label{experiment_set_up} 
\subsubsection{Experimental sample}
The efficiencies calculated with the MaLBEC were compared to the efficiencies calculated with one of the traditional methods, MCNP, using a steel sample from the Joint European Torus (JET) hall. The sample was activated by fusion neutrons during deuterium-deuterium and deuterium-tritium campaigns. The sample was measured with a HPGe gamma spectrometry detector in the RADLab. The experimental set-up of the sample and the resulting spectrum are shown in figure~\ref{fig:efficiency}. Gamma rays identified in figure~\ref{fig:efficiency} from manganese-54 (Mn-54) and Co-60 are due to the \(^{59}\mathrm{Co}(n,\gamma)^{60}\mathrm{Co}\) and \(^{54}\mathrm{Fe}(n,p)^{54}\mathrm{Mn}\) reactions in the steel. The information gathered from this sample (and others like it) provides vital insights for fusion research. For example, the gamma information can be used to infer the number of incident neutrons during irradiation \cite{ross_unfolding}, which in turn provides information on fusion power. As such, this was chosen to demonstrate the effectiveness of the MaLBEC.   

\subsubsection{Efficiency calculation results}
The efficiency results for the experimental measurement were calculated with MCNP (as the traditional/baseline method) and the MaLBEC. Four metrics were used to compare the two methods. 

\begin{figure}[H]
    \centering
    \includegraphics[width=0.5\textwidth]{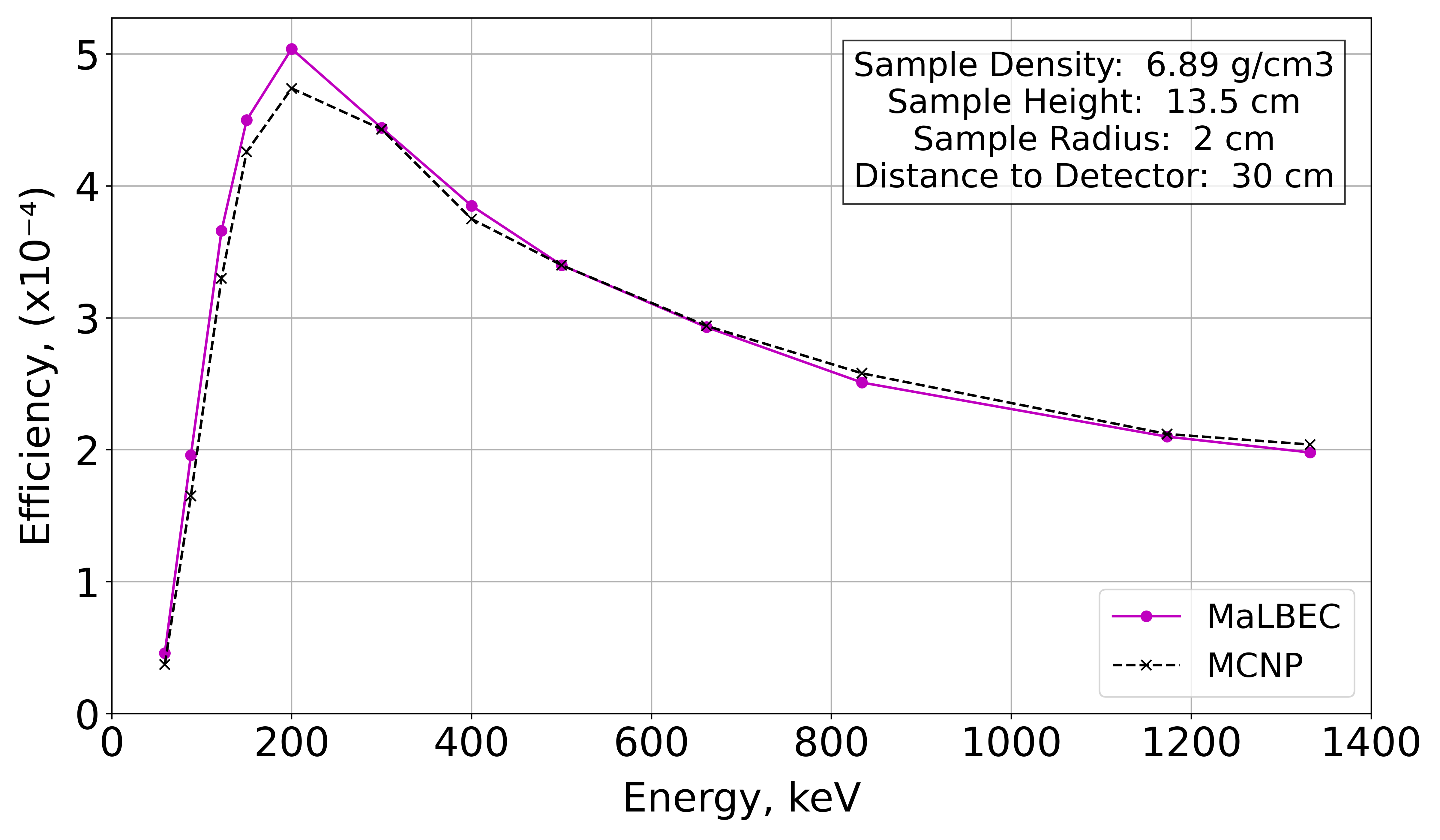}
    \caption{Comparison of the efficiency curves produced by MCNP (as a commonly used method, in black) and by the MaLBEC (pink), for the steel sample gamma spectrometry measurement}
    \label{fig:experiment_results}
\end{figure}

The first metric was a comparison of the efficiency values per energy, based on the geometry and positioning of the steel sample. To compare the results from MCNP and the MaLBEC, the percentage difference from equation~\ref{percent_dif} was used to determine how close the MaLBEC results were to the MCNP method. The percentage difference on average was 5\%, and all were less than 20\% or lower (table~\ref{tab:percent_difference}), which is acceptable in most cases in an industrial setting. The efficiency value at 834~keV was interpolated from the 661~keV and 1173~keV gamma ray energy lines, as this value was not included in the training data for the MaLBEC. The efficiency curves for both are plotted in figure~\ref{fig:experiment_results}, further illustrating how well the MaLBEC performs compared MCNP due to the closeness of the curves. The higher percentage differences at low gamma ray energies (e.g. 59–100~keV) are likely due to increased sensitivity to material attenuation, modelling precision, and statistical uncertainty in simulations. At higher energies, these effects diminish, resulting in closer agreement between the two methods.

\begin{table}[h]
\centering
\caption{Percentage difference between MCNP and the MaLBEC efficiency values, for the 11 gamma ray energy values.}
\label{tab:percent_difference}
\begin{tabular}{cccc}
\toprule
Energy, keV & MCNP        & MaLBEC & \%diff  \\ \hline
59      &  3.72E-05       & 4.60E-05 & 23.6    \\ 
88      &  1.65E-04       & 1.96E-04 & 19.0    \\ 
122     &  3.30E-04       & 3.66E-04 & 11.0    \\ 
150     &  4.26E-04       & 4.50E-04 & 5.6     \\ 
200     &  4.74E-04       & 5.04E-04 & 6.4     \\ 
300     &  4.43E-04       & 4.44E-04 & 0.3     \\ 
400     &  3.75E-04       & 3.85E-04 & 2.5     \\ 
500     &  3.40E-04       & 3.40E-04 & 0.1     \\ 
661     &  2.94E-04       & 2.93E-04 & -0.5     \\ 
883     &  2.58E-04       & 2.51E-04 & -2.5     \\ 
1173    &  2.12E-04       & 2.10E-04 & -0.8     \\ 
1332    &  2.04E-04       & 1.98E-04 & -2.7     \\ 
\botrule
\end{tabular}
\end{table}

The second metric was a comparison of the derived activities of the steel sample, using the relevant efficiency values from MCNP and the MaLBEC in equation~\ref{eq1}. Equation~\ref{percent_dif} was used to determine the difference between the activity derived using the efficiency from MCNP and the MaLBEC. The results are shown in table~\ref{tab:activity_comp}, and the difference is less than 3\% for the two nuclides identified in the spectrum.

\begin{table}[h]
\centering
\caption{Percentage difference for the steel sample between using the efficiency from MCNP and the MaLBEC to calculate the activity of the nuclides identified.}
\label{tab:activity_comp}
\begin{tabular}{ccccc}
\toprule
\multicolumn{1}{c}{\begin{tabular}[c]{@{}c@{}}Energy, \\ keV \end{tabular}} &
Nuclide &
\multicolumn{1}{c}{\begin{tabular}[c]{@{}c@{}}MCNP \\ activity, Bq \end{tabular}} &
\multicolumn{1}{c}{\begin{tabular}[c]{@{}c@{}}MaLBEC \\ activity, Bq \end{tabular}} &
\multicolumn{1}{c}{\begin{tabular}[c]{@{}c@{}}\%diff \end{tabular}} 
\\ \hline
834   & Mn-54    & 1.01E+03   &  1.03E+03 & 2.5 \\
1173  & Co-60    & 2.16E+03   &  2.17E+03 & 0.8 \\
1332  & Co-60    & 2.27E+03   & 2.33E+03  &  2.7 \\
\botrule
\end{tabular}
\end{table}

The third metric was a comparison of the computational speed to get efficiency values, from the point of submission, but not including the model build time as this is more subjective. For MCNP, the computational time was taken for each of the 11 gamma ray energies per geometry and summed. The percentage difference from equation~\ref{percent_dif} was used again in this comparison to compare the time for each method to produce all 11 efficiency results. The results show a 99.96\% decrease in computational time for the MaLBEC, which showcases the exceptional speed in which the MaLBEC produces results for all 11 efficiencies - less than 1~second, compared to 2627~seconds (44~minutes) for MCNP.  

The final metric was designed to compare the usability of each method, by comparing the number of code edits required to produce efficiency values for all 11 gamma ray energies of a new cylindrical sample. This metric depends on how the MCNP file was created, but it has been standardised for this work. For the MCNP files, 23 edits were required to set up the sample dimensions and position, and a further three edits for each of the 11 gamma ray energies. Therefore, to attain a full efficiency curve for a new sample, 56 edits were required. In contrast, only four edits were required for the MaLBEC, once initial training had been performed. This is a significant simplification and vastly improves usability. 

\section{Discussion}\label{sec4}
This work introduces a novel algorithm, the machine learning based efficiency calculator (MaLBEC), which utilises state-of-the-art machine learning methods to calculate the absolute efficiencies at 11 gamma ray energies for a sample, based on just four inputs (compared to 56 inputs for MCNP). The MaLBEC produced efficiency results for a HPGe gamma spectrometry detector and a typical fusion sample measurement, that were within 5\% of a widely used efficiency calculation method, MCNP. Compared to the traditional method using the UKAEA high-performance computing cluster, which took 45~minutes to produce results, the MaLBEC approach achieved the same outcome in just 1~second, which represents a reduction in computation time of over 99.96\%. When used to determine the activity of nuclides in a fusion sample, the MaLBEC produced results that were within 3\% of the traditional MCNP method. This demonstrates that the MaLBEC provided a robust alternative method to determining the efficiency for experimental measurements. It was demonstrated that the MaLBEC performed well and did not overfit the training data, but showed good generalisation to new, unseen data.

The dramatic improvement in usability and speed, combined with high accuracy compared to a traditional method, enables simple and fast processing, significantly enhancing the efficiency calculation process. The MaLBEC requires an input of only four values (sample density, sample height, sample radius, and detector-to-sample distance), making the algorithm simple to use for even untrained users. This lowers the barrier for all users of gamma spectrometry to perform efficiency calculations independently from costly, complex, or inaccessible traditional methods. This has the potential to improve the fusion diagnostic process, by simplifying and accelerating the efficiency calculation process, enabling rapid analysis of samples relevant to fusion calculations.

Future developments of the MaLBEC would include diversifying the sample geometries for which it can be used, including more shapes, sizes, and materials. A second development would be to include varying detector crystal sizes, as this would make the algorithm detector agnostic and would dramatically increase its usability across different systems. Another development would be to expand the outputs of the algorithm, to enable the user to define the energy at which they require the efficiency value for, further simplifying the process and reducing potential inaccuracies with interpolation. To achieve this, training data across a continuous spectrum of gamma ray energies would be required, rather than at the 11 discrete gamma ray energies considered here. Finally, due to the simplicity of this tool, created in the Python programming language, future development would look to build the MaLBEC into other laboratory automation tools, further simplifying and improving the entire gamma spectrometry process.


\backmatter

\section*{Declarations}
\subsection*{Funding}
This work has been part-funded by the EPSRC Energy Programme (grant number EP/W006839/1) and the UK STFC (grant number ST/Y000331/1). To obtain further information on the data and models underlying this paper please contact PublicationsManager@ukaea.uk.

\subsection*{Rights retention statement}
For the purpose of open access, the author has applied a Creative Commons Attribution (CC BY) licence to any Author Accepted Manuscript version arising from this submission.

\subsection*{Conflict of interest}
The authors have no competing interests to declare that are relevant to the content of this article.

\subsection*{Data Availability}
All data generated or analysed in this work are available on request.

\bibliography{bibliography}

\end{document}